\documentclass[entropy,article,accept,oneauthors,latex,12pt,a4paper]{mdpi}
\lastpage{x}
\doinum{10.3390/------}
\pubvolume{xx}
\pubyear{2016}
\history{Received: xx / Accepted: xx / Published: xx}

\usepackage{graphicx}
\usepackage{amsmath}

\newcommand{\psia}{a}

\newcommand{\myskip}[1]{}

\newcommand{\minfty}{{-\infty}}

\newcommand{\vGam}{{\bf \Gamma}}

\newcommand{\upeen}{{(1)}}
\newcommand{\vG}{{\bf G}}

\newcommand{\va}{{\bf a}}

\newcommand{\ve}{{\bf e}}
\newcommand{\vf}{{\bf f}}
\newcommand{\vh}{{\bf h}}

\newcommand{\vs}{{\bf s}}
\newcommand{\vu}{{\bf u}}
\newcommand{\vv}{{\bf v}}
\newcommand{\vw}{{\bf w}}

\newcommand{\ccdot}{\hspace{-0.5mm}\cdot\hspace{-0.5mm}}

\newcommand{\vB}{{\bf B}}
\newcommand{\vE}{{\bf E}}

\newcommand{\vC}{{\bf C}}

\newcommand{\vJ}{{\bf J}}

\newcommand{\vL}{{\bf L}}

\newcommand{\vS}{{\bf S}}
\newcommand{\vR}{{\bf R}}

\newcommand{\cE}{{\cal E}}

\renewcommand{\vf}{{\bf f}}

\newcommand{\vp}{{\bf p}}
\newcommand{\vr}{{\bf r}}

\newcommand{\p}{{\partial}}
\newcommand{\veen}{{\bf 1}}
\newcommand{\vnul}{{\bf 0}}

\renewcommand{\p}{\partial}

\newcommand{\BEQ}{\begin{eqnarray}}
\newcommand{\EEQ}{\end{eqnarray}}
\newcommand{\BEA}{\begin{eqnarray}}
\newcommand{\EEA}{\end{eqnarray}}

\newcommand{\nn}{\nonumber }
\renewcommand{\d}{{{\rm d}}}
\newcommand{\eps}{\varepsilon}

\newcommand{\eff}{{\rm eff}}
\newcommand{\half}{\frac{1}{2}}

  \renewcommand{\thesection}{\arabic{section}}
   \renewcommand{\theequation}{\thesection.\arabic{equation}}
     \renewcommand{\thesection}{\arabic{section}}
 \renewcommand{\thesubsection}{\thesection\arabic{subsection}}
    \renewcommand{\theequation}{\thesection\arabic{equation}}

\Title{On the stability of classical orbits of the hydrogen ground state in Stochastic Electrodynamics}

\Author{Theodorus Maria Nieuwenhuizen$^{1,2}$}

 \address{%
$^1$ Institute for Theoretical Physics, 
P.O. Box 94485, 1098 XH, Amsterdam, the Netherlands \\
$^2$International Institute of Physics, UFRG,  Lagoa Nova, Natal - RN, 59064-741, Brazil}

\corres{e-mail: t.m.nieuwenhuizen\@ uva.nl;  telephone +31-20-525-6332}

\abstract{de la Pe\~na 1980 and Puthoff 1987 show that circular orbits in the hydrogen problem of Stochastic Electrodynamics are stable. 
Though the Cole-Zou 2003 simulations support the stability, our recent numerics always lead to self-ionisation. 
Here the de la Pe\~na-Puthoff argument is extended to elliptic orbits. 
For very eccentric orbits with energy close to zero and angular momentum below some not-small value, 
there is on the average a net gain in energy for each revolution, which explains the self-ionisation.
Next, an $1/r^2$ potential is added, which could stem from a dipolar deformation of the nuclear charge by the electron at its moving position.
This shape retains the analytical solvability. When it is enough repulsive,
the ground state of this modified hydrogen problem is predicted to be stable.
The same conclusions hold for positronium.
}

\keyword{Stochastic Electrodynamics; hydrogen ground state; stability criterion}

\PACS{11.10 \and 05.20 \and 05.30 \and 03.65}

\begin{document}

\section{Introduction}
\label{intro}

Stochastic Electrodynamics (SED) is a subquantum theory that considers the quantum vacuum as a true physical vacuum with its zero-point modes 
being physical electromagnetic modes, see the books \cite{cetto1996quantum,delapena2014emerging}. 
By construction, the vacuum is filled with classical electromagnetic modes, each carrying an energy $\half\hbar\omega$,
where Planck's constant is a constant of nature characterising the energy stored in these modes. The task is then to show that SED explains 
all quantum behaviour of matter at the statistical level. This has been argued to occur on general grounds \cite{cetto1996quantum,delapena2014emerging}.
SED is not ruled out by Bell's theorem, since that has an irreparable fatality, the contextually loophole \cite{nieuwenhuizen2011contextuality}.

Testing the working of SED in special cases is needed to get trust in the theory. Since harmonic oscillators perform well, several phenemona
are explained: van de Waals forces, the logarithm of the Lamb shift between the hydrogen $1s$ and $2p$ states, the Casimir effect, the Unruh effect.
A natural next case is the hydrogen ground state. de la Pe\~na \cite{delapena1980introduccion} and Puthoff  \cite{puthoff1987ground,puthoff2012quantum}
demonstrate that circular orbits are lead to stable atom: they do not fall onto the centre and they do not self-ionise.
The SED theory has been tested numerically on the hydrogen ground state in a 2-$d$ approximation where the orbit remains in its initial plane.
Cole and Zou 2003 observed an encouraging agreement with the result from the quantum ground state wave function \cite{cole2003quantum}. 
New simulations have been carried out recently by our team \cite{nieuwenhuizen2015simulation1} in a $3d$ approach, benefiting from  
a decade of progress in computational and programming power and employing a few analytical tricks to make the problem tractable in $3d$.
However, it was observed that in all runs and all attempts to model the system, there occurred self-ionisation.
The latter fact stems with a theoretical prediction \cite{claverie1982nonrecurrence}. Boyer puts forward that the problem may lie
in relativistic effects \cite{boyer2015classical},  but taking into account relativistic corrections in our computer code did not change 
much and in particular did not cure the self-ionisation \cite{nieuwenhuizen2015simulation2}.

Faced with the renewed finding of self-ionisation, we search a theoretical underpinning for it and
reconsidering the stability argument of de la Pe\~na and  Puthoff \cite{delapena1980introduccion,puthoff1987ground}, we extend it to include eccentricity.
We derive the gain and loss terms, averaged over the stochastic force and averaged over a full period in section 2. 
In order to study a more general problem with possibly a stable state,
we extend in section 3 the problem with an attractive or repulsive $1/r^2$ potential.
We close with a discussion. In an appendix we consider the case where both masses are finite, such as the positronium.

\renewcommand{\thesection}{\arabic{section}}
\section{On the stability of the hydrogen ground state}
\setcounter{equation}{0}\setcounter{figure}{0} 
 \renewcommand{\thesection}{\arabic{section}.}

\newcommand{\phinul}{\chi}

The nonrelativistic equation of motion of an electron with charge $-e$ around an ion of charge $Ze$ can be written in a dimensionless form
where distances are measured in terms of the Bohr radius $\hbar/(\alpha Z m_ec)$ and times in the Bohr time $\hbar/\alpha^2Z^2m_ec^2$ 
\cite{nieuwenhuizen2015simulation1},

\BEQ
\label{rddotnr=}
\ddot\vr=&-&\frac{\vr}{r^3}-\beta \vE+\beta^2\dddot\vr, \qquad \beta=\sqrt{\frac{2}{3}}\alpha^{3/2}Z,
\EEQ
where $\alpha=1/137$ the fine structure constant and $\vE(t)$ the fluctuating electromagnetic field that lies at the heart of SED.
The term $\beta^2\dddot\vr$ arises from damping.
We consider a time window during which the orbit remains 
basically unperturbed and in a plane, which is meaningful because of the small value of $\beta\sim Z/2000$.
The correlator of the $\vE$ field reads in these units, with a factor $\sqrt{3/2}$ absorbed in $\vE$ \cite{nieuwenhuizen2015simulation1}, 

\BEQ
\vC_E(t-s)= \langle \vE(t)\vE(s)\rangle =\veen \times C_E(t-s),\quad C_E(t-s)=\frac{6}{\pi}\Re\frac{1}{(t-s+i\tau_c)^4}.
\EEQ
and it is expected that the cutoff $\tau_c\sim \alpha^2 Z^2$, corresponding in physical units to the Compton time 
$\hbar/m_ec^2$, can be taken to 0 at the end.

Let at the initial time $t_0$ the orbit lie in the $x-y$ plane, with $\vr=r\ve_1(t)$ where $\ve_1(t)=(\cos\phi(t),\sin\phi(t),0 $) 
and likewise  $\ve_2(t)=(-\sin\phi(t),\cos\phi(t),0 $). The third unit vector is $\ve_3=\vL/L=(0,0,1)$.

The unperturbed orbit  can be coded as
 
\BEQ r^{(0)}(t)=\frac{L^2}{1+\eps\cos[\phi(t)-\chi]},\qquad \dot\phi(t)=\frac{L}{r^2(t)}=\frac{\{1+\eps\cos[\phi(t)-\chi]\}^2}{L^3},\quad
\EEQ
so that $\dot r^{(0)}(t)=(\eps/L)\sin[\phi(t)-\chi]$.
Here $L$, $\eps$ and $\chi$ are the angular momentum,  eccentricity and angle between long axis and the $x$-axis,
all conserved in the Kepler problem, that is, in the absence of stochastic fields and damping.
The energy of this orbit is $\cE=\half\dot\vr^2-1/r=-\half k^2$ . The parameters $\eps$, $k$ and $L$ are related by

\BEQ
\kappa\equiv \sqrt{1-\eps^2},\qquad \lambda\equiv kL,\qquad \kappa=\lambda.
\EEQ
Let $\psia $ code the angle $\phi$ according to 

\BEQ\label{phi2a1}
\sin(\phi-\phinul )=\frac{\kappa\,\sin\psia}{1-\eps\cos\psia},\quad 
\cos(\phi-\phinul )=\frac{\cos\psia-\eps}{1-\eps\cos\psia},\quad 
\EEQ
with inverse 

\BEQ
\cos\psia=\frac{\cos(\phi-\phinul )+\eps}{1+\eps\cos(\phi-\phinul )},\quad
\sin\psia=\frac{\kappa\sin(\phi-\phinul )}{1+\eps\cos(\phi-\phinul )} .\qquad
\tan\frac{\phi-\chi}{2}=\sqrt{\frac{1+\eps}{1-\eps}}\tan\frac{a}{2}.
\EEQ

Here the trace led to inner products of  the $\ve$'s with the connection, following from Eq. (\ref{phi2a1}),

\BEQ \label{sinxycosxy}
\cos[\phi(t)-\phi(s)]&=&
\frac{\cos(a-b)-\eps(\cos a+\cos b)+\eps^2(1-\sin a\sin b)}{(1-\eps\cos a)(1-\eps\cos b)},\qquad \nn\\
\sin[\phi(t)-\phi(s)]& =& 
\kappa \frac{\sin(a-b)- \eps (\sin a - \sin b)}{(1 - \eps \cos a)(1 - \eps \cos b)},
\EEQ

The orbit may be written in the parametric form

\BEQ
r^{(0)}=\frac{1-\eps\cos\psia}{k^2},\quad 
t=\frac{\psia-\eps\sin\psia}{k^3},\quad
\dot r^{(0)}=\frac{k\eps\sin\psia}{1-\eps\cos\psia}.
\EEQ
The angle, starting with $\phi=0$ at $t=0$, evolves as

\BEQ\label{phiamu1=}
\phi(t)=\phi_a=2\arctan\frac{1+\eps}{\kappa}\tan\frac{a}{2}=2\overline{\arctan}\frac{1+\eps}{\kappa}\tan\frac{a}{2}+2\pi \left\lfloor \frac{\pi+a}{2\pi}\right \rfloor,
\EEQ
where $\overline{\arctan}$ takes values between $-\pi/2$ and $\pi/2$ and is discontinuous at $a=(2n+1)\pi$ for integer $n$; 
the floor function $\lfloor x \rfloor$ delivers the largest integer below $x$.

\subsection{Perturbations around the Kepler orbit}

The energy changes due to the stochastic field and due to radiation. 
Due to the small value of $\alpha$ these effects can be calculated separately.
In a perturbation expansion to order $\beta^2$ the  rate of energy change due to the stochastic field 
follows from

\BEQ
\vr(t)=\vr^{(0)}(t)
+\beta\vr^{(1)}(t)+{\cal O}(\beta^2)
\EEQ
as

\BEQ
\dot \cE_{\rm field}(t)=-\beta\vE(t)\ccdot[\dot\vr^{(0)}(t)+\beta \dot\vr^{(1)}(t)].
\EEQ
With the leading term vanishing on the average, there remains the statistically averaged 
energy gain from the field

\BEQ
\langle \dot \cE_{\rm field}(t)\rangle =\beta^2 \langle \vE(t)\ccdot \dot \vr^{(1)}(t)\rangle. 
\EEQ
We shall suppose that the stochastic fields and the damping start to act at time $t_0$, which is later taken to $\minfty$.
The solution for the perturbation $\vr^\upeen$ has the form

\BEQ \label{r1G}
\vr^{(1)}(t)=-\int_{t_0}^t\d s\, \vG(t,s)\ccdot\vE(s).
\EEQ
with $\vG$ the Green's function, to be constructed.
Hence the average effect by the field, for times during which the Kepler orbit is not disturbed much, reads 
in terms of $\dot\vG(t,s)\equiv\p_t\vG(t,s)$,

\BEQ \label{Edotfield}
\langle \dot \cE_{\rm field}(t)\rangle=
\beta^2\int_{t_0}^t\d s \, {\rm tr}_{3d}[\dot \vG(t,s)\ccdot \vC_E(t-s)] 
=\beta^2\int_{t_0}^t\d s \, {\rm tr}_{3d}[\dot \vG(t,s)]\,C_E(t-s).
\EEQ

\myskip{
\subsubsection{Wiggling in the plane of the orbit}

We solve the Kepler dynamics in the rotating frame where the field has components
$E_{1,2}(t)=\ve_{1,2}(t)\ccdot\vE(t)$, and the orbit is decomposed on the rotating eigenvectors $\ve_{1,2}$ and the fixed perpendicular vector $\ve_3$,

\BEQ
\vr^\upeen (t)=r_{1}(t)\ve_1(t)+r_{2}(t)\ve_2(t)+z_1(t)\ve_3.
\EEQ
This brings the equations
$\ddot\vr^\upeen=\nabla\vf\cdot\vr^\upeen+\vE$, which reads in the rotating frame

\BEQ\label{r1r2eq}
\ddot r_{1}- r_{1}\dot\phi^2-2\dot r_{2}\dot\phi-\ddot\phi r_{2}
-\frac{2r_{1}}{r^3}&=&E_1 \nn\\
2\dot r_{1}\dot\phi+\ddot\phi r_{1}+\ddot r_{2}- r_{2}\dot\phi^2
+\frac{r_{2}}{r^3}&=&E_2 \nn\\
\ddot z_1+\frac{z_1}{r^3}&=&E_3
\EEQ
where $\dot\phi=L/r^2$ and $r$ are the zero order solutions. 
}

\subsection{Perturbations of the orbit}

We solve the perturbed Kepler dynamics in the rotating frame where 
the orbit is decomposed on the rotating eigenvectors $\ve_{1,2}$ and the fixed perpendicular vector $\ve_3$.
The field then has components $E_{1,2,3}(t)=\ve_{1,2,3}(t)\ccdot\vE(t)$.
The equation $\ddot\vr_1+\nabla\nabla V\cdot\vr_1=\vE$ reads on the comoving  basis

\BEQ\label{r1r2eq}
\ddot r_{1}- r_{1}\dot\phi^2-2\dot r_{2}\frac{L}{r^2}+2\frac{\dot rL}{r^3}r_{2}-\frac{2}{r^3}r_1
&=&E_1 \nn\\
\ddot r_{2}- r_{2}\dot\phi^2+2\dot r_{1}\dot\phi+\ddot\phi r_{1}+\frac{1}{r^3}r_2&=&E_2 \\
\ddot z_1+\frac{1}{r^3}z_1&=&E_3  \nn
\EEQ
Let us search the homogeneous solutions $\vh=(r_1,r_2,z_1)$.
The first is $\vh^{(1)}(t)\sim \dot\vr(t)$. Another one, $\vh^{(3)}$,  has $r_1=0$, $r_2\sim r(t)\sim\rho_a$.  
Two further in-plane solutions $\vh^{(2,4)}$ are more intricate but can be constructed analytically. $\vh^{(2)}$ contains oscillations around a term linear in $t$. 
For $z_1$ one finds the solutions $r\cos\phi$ and $r\sin\phi$ from components of the orbit on the non-rotating basis. 
We present these six homogeneous solutions as

\BEQ
\vh^{(1)}(t)&=&\frac{1}{\rho_a}\left(\begin{array}{c}  \eps\sin a  \\  \kappa \\0  \end{array}\right),  \nn \qquad  
\vh^{(2)}(t)=\frac{1}{\rho_a}\left(\begin{array}{c} \kappa^2( \cos a -\eps) - 2\eps\rho_a^2 \\ -\kappa (\kappa^2+\rho_a)\sin a  \\ 0 \end{array}\right) 
+3\eps\tau_a\vh^{(1)},  \qquad \\
\vh^{(3)}(t)&=&\rho_a\left(\begin{array}{c} 0 \\  1   \\ 0 \end{array}\right),  \qquad  \qquad \,
\vh^{(4)}(t)=\frac{1}{\rho_a}\left(\begin{array}{c} \kappa(\eps-\cos a)  \\ 
(\kappa^2+\rho_a)\sin a \\ 0  \end{array}\right),    \\
\vh^{(5)}(t)&=&\rho_a\sin\phi_a \left(\begin{array}{c}0 \\ 0 \\1\end{array}\right), \qquad \!
\vh^{(6)}(t)=\frac{\eps}{\kappa}\rho_a\cos\phi_a\left(\begin{array}{c}0 \\ 0 \\1\end{array}\right),\qquad   \nn
\EEQ
where $\tau_a=a-\eps\sin a=k^3t$  and  $ \rho_a= \d\tau_a/d a=1-\eps\cos a$.

\myskip{
\BEQ
\vh^{(1)}(t)&=&\frac{1}{\rho_a}\left(\begin{array}{c}  \eps\sin a  \\  \kappa \\0  \end{array}\right),  \nn \qquad  
\vh^{(2)}(t)=\frac{1}{\kappa}\left(\begin{array}{c} 0 \\  \rho_a   \\ 0 \end{array}\right),  
\\
\vh^{(3)}(t)&=&\frac{1}{\rho_a}\left(\begin{array}{c} \kappa^2( \cos a -\eps) - 2\eps\rho_a^2 \\ -\kappa (\kappa^2+\rho_a)\sin a  \\ 0
\end{array}\right) +3\eps\tau_a\vh^{(1)},  \qquad 
\vh^{(4)}(t)=\frac{\kappa}{\rho_a}\left(\begin{array}{c} \kappa(\eps-\cos a)  \\ 
(\kappa^2+\rho_a)\sin a \\ 0  \end{array}\right),
\nn\\
\vh^{(5)}(t)&=&\rho_a\cos\phi_a\left(\begin{array}{c}0 \\ 0 \\1\end{array}\right),\qquad 
\vh^{(6)}(t)=\frac{\eps}{\kappa}\rho_a\sin\phi_a \left(\begin{array}{c}0 \\ 0 \\1\end{array}\right).
\EEQ
where $\tau_a=a-\eps\sin a=k^3t$  and  $ \rho_a= \d\tau_a/d a=1-\eps\cos a$.
}

The solution (\ref{r1G}) then involves the Greens function $\vG$, which on the rotating basis is denoted as ${\bf \Gamma}$,

\BEQ
G_{ij}(t,s)
=\sum_{k,l=1}^3e_i^{(k)}(t)\Gamma_{kl}(t,s)e_j^{(l)}(s)
\EEQ
The conditions $\vG(t,t)=\vnul$ and $\dot\vG(t,t)=\veen$ imply that also $\vGam(t,t)=\vnul$ and $\dot\vGam(t,t)=\veen$.
Some effort leads to the elegant exact solution

\BEQ \label{Gam=}
\vGam(t,s)=\sum_{i=1,3,5}
\frac{\vh^{(i)}(t)\vh^{(i+1)}(s)
-\vh^{(i+1)}(t)\vh^{(i)}(s)}{\eps k^3}  .
\EEQ
\myskip{
\BEQ \label{Gam=}
\vGam(t,s)=
\frac{
\vh^{(1)}(t)\vh^{(3)}(s)
-\vh^{(3)}(t)\vh^{(1)}(s)
+\vh^{(2)}(t)\vh^{(4)}(s)
-\vh^{(4)}(t)\vh^{(2)}(s)
+\vh^{(5)}(t)\vh^{(6)}(s)
-\vh^{(6)}(t)\vh^{(5)}(s)}{\eps k^3}
\EEQ
\BEQ \label{Gam=}
\vGam(t,s)=
\frac{1}{\eps k^3}\left[
\vh^{(1)}(t)\vh^{(3)}(s)
-\vh^{(3)}(t)\vh^{(1)}(s)
+\vh^{(2)}(t)\vh^{(4)}(s)
-\vh^{(4)}(t)\vh^{(2)}(s)
\right] +\Gamma_{33}\ve_3\ve_3,
\EEQ}
$\Gamma_{33}$ comes from the transverse fluctuations, i.e., from $\vh^{(5,6)}$, with the result

\BEQ
G_{33}=\Gamma_{33}
=\frac{(1-\eps\cos a)(1-\eps\cos b)}{\kappa k^3}\sin \phi_{ab}
=\frac{\rho_a\rho_b}{\kappa k^3}\sin \phi_{ab}
\EEQ
Here $\phi_{ab}=\phi_a-\phi_b$ with the rotation angle from (\ref{phiamu1=}).
It is easily verified that ${\bf W}(t)=\dot\vGam(t,s)|_{s\to t}$ is a matrix valued Wronskian, that is, it is a constant matrix, indeed equal to  $\veen$.

\subsection{Statistical rate of energy change from the stochastic field}

The field contribution to the energy change reads from (\ref{Edotfield})  (taking from now on $t_0\to\minfty$)

\BEQ \label{dotE=}
\langle\dot\cE\rangle_{\rm field}=\beta^2\int_{\minfty}^t\d s\, \dot g(t,s)C_E(t-s), \qquad 
 \dot g(t,s)\equiv {\rm tr}_{3d}\dot\vG(t,s)-3.
 \EEQ
where

\BEQ
{\rm tr}_{3d} \vG=(\Gamma_{11}+\Gamma_{22})\cos(\phi_t-\phi_s)+ (\Gamma_{12}-\Gamma_{21})\sin(\phi_t-\phi_s)+G_{33}.
\EEQ
and the dot denotes differentiation to $t$.
The subtracted term in (\ref{dotE=}), cancelling $\dot\vG(t,t)=\veen$, does not contribute to the integral.

From the result

\BEQ
\dot G_{33}(t,s)&=&1+\frac{\cos({a-b})-1}{1-\eps \cos a},
\EEQ
an expansion in $b$ near $b=a$ yields, when expressed as a series in $t-s$,

\BEQ
\dot G_{33}(t,s)=1-
\frac{k^6(t-s)^2}{2(1-\eps\cos a)^3}-\frac{\eps\sin a \,k^9(t-s)^3}{2(1-\eps \cos a)^5}+{\cal O}[(t-s)^4],
\EEQ
whereas the in-plane part behaves as

\BEQ
{\rm tr}_{2d} \dot\vG(t,s)=2+\frac{k^6(t-s)^2}{2(1-\eps \cos a)^3}+\frac{\eps\sin a \,k^9(t-s)^3}{2(1-\eps \cos a)^5}+{\cal O}[(t-s)^4].
\EEQ
As seen in Eq. (\ref{tau20}) below, the second and third order terms are potentially dangerous for the convergence of the integral at $s=t$, 
but they are exactly opposite, so that $g(t,s)={\cal O}[(t-s)^4]$ for $s\to t$.
This behaviour assures that in the limit $\tau_c\to 0$ the integral

\BEQ\label{tau20}
\lim_{\tau_c\to0}\Re\int_{\minfty}^t\d s\,\frac{g(t,s)}{(t-s+i\tau_c)^4}
=\int_{\minfty}^t\d s\,\frac{\dot g(t,s)}{(t-s)^4},
\EEQ
behaves well, as it should, since no subtle short-time effects are expected for the energy absorbed from the field.
The combined effect takes the compact form 

\BEQ
g(t,s)\equiv {\rm tr}_{3d}\vG(t,s) & =&
 \frac{A(a,b) 
 +B(a,b)(a-b-\eps\sin a+\eps \sin b)}{(1-\eps\cos a)(1-\eps\cos b)k^3}, 
 \EEQ
 with antisymmetric $A$ and symmetric $B$,
 

\BEQ
A(a,b)&=&5 \sin (a - b) + \half \sin 2 (a - b)
+     \frac{3}{2}\eps^2  [\sin 2 a-\sin 2b + 2 \sin( a - b)] \nn\\
&&-  2\eps[3 (\sin a-\sin b) + \sin(a - 2 b) + \sin (2 a - b)]  ,
          \\
B(a,b)&=&-3\cos(a-b)+3\eps^2\cos a\cos b . \nn
 \EEQ

We express the energy gain from the field, averaged over one period, as
\BEQ \label{Edotfieldd0}
\langle\dot\cE_{\rm field}\rangle
=\beta^2\frac{k^9}{\kappa^6}(1+\half\eps^2)f(\kappa),
\EEQ
where

\BEQ
f(\kappa)=\frac{6\kappa^6}{\pi^2(3-\kappa^2)}\int_{-\pi}^\pi\d a\int_\minfty^a\d b\frac{(1-\eps \cos a)(1-\eps\cos b)
\dot g(t,s)}{(a-b-\eps\sin a+\eps \sin b)^4}.
\EEQ
It is a quite smooth function, ending for circular orbits ($\eps=0$, $\kappa=1$)
 at $f(1)=\frac{1}{2}$. The maximal value at $\eps=1$, $\kappa=0$ is obtained by scaling $a\to\kappa u$,
$b\to\kappa v$, which yields for $\kappa\to0$

\BEQ \label{f(1)-int}
f(0)=\frac{2^3 3^4 }{5\pi^2}\int_\minfty^\infty\d u\int_\minfty^u\d v\,
\frac{5 + 3 u^2 + 8 u v - v^2+ 4 u^3 v  + u^2 v^2}{ (1 +    u^2)^2 (3 + u^2 + u v + v^2)^4}.
\EEQ
To evaluate this, one may analytically perform the $v$-integral, which leads to several terms odd and one even in $u$,
the latter resulting in a result not far above $f(1)=\half$:


\BEQ
f(0)=\frac{16}{5\pi \sqrt{3}}=0.588084.
\EEQ

\subsubsection{Radiative energy loss}

The loss terms produce a rate of energy loss averaged over one orbit of period $P=2\pi/k^3$ equal to

\BEQ \label{Eradgen}
\langle\dot\cE_{\rm rad}\rangle &=& 
\beta^2\langle\dddot\vr\cdot\dot\vr \rangle=-\beta^2\langle\ddot\vr^2 \rangle =-\beta^2\langle\vf^2 \rangle \nn \\
&=&-\beta^2\langle\frac{1}{r^4}\rangle =-\frac{\beta^2}{P}\int_{0}^P\d t\frac{1}{r^4},
\EEQ
where,  in the averaging over a full period, we could omit a total derivative.
Using $\d t=\d\phi/\dot\phi=r^2\d\phi/L$ and $\kappa=kL$ this becomes

\BEQ
\langle\dot\cE_{\rm rad}\rangle=-\frac{\beta^2k^3}{2\pi L^5}\int_{0}^{2\pi} \d \phi\,(1+\eps\cos\phi)^2
=-\beta^2k^8\frac{3-\kappa^2}{2\kappa^5}
\EEQ

Combining gain and loss terms leads to the total energy change 
\BEQ \label{Etotdot}
\langle\dot\cE_{\rm tot}\rangle
=\beta^2\frac{k^8}{2\kappa^6}(2+\eps^2)[\,kf(\kappa)-\kappa].
\EEQ
For spherical orbits ($\eps=0$, $\kappa=1$) there is the explicit result, first derived qualitatively by
de la Pe\~na \cite{delapena1980introduccion}  and  Puthoff \cite{puthoff1987ground}

\BEQ
\langle\dot\cE_{\rm tot}\rangle
=\beta^2k^8(\frac{k}{2}-1).
\EEQ
It exhibits stability: for large negative energy $\cE=-\half k^2$ ($k\gg 1$) the energy gain is positive on the average, 
preventing collapse on the nucleus.
On the other hand, when $\cE$ is small negative, its average change is negative, preventing $\cE\ge0$, i.e., self-ionisation.

But with $\kappa=kL\to0$ at fixed $L$, Eq.  (\ref{Etotdot}) yields

\BEQ
\langle\dot\cE_{\rm tot}\rangle
=\beta^2\frac{3k^3}{2L^6}[f(0)-L].
\EEQ
Per period in the limit $k\to0$ at fixed $L$ this implies a fixed amount,

\BEQ \label{DeltaEd0}
\Delta\langle\cE_{\rm tot}\rangle
=\beta^2\frac{3\pi}{L^6}[f(0)-L].
\EEQ
In words: each revolution produces, on the average, a finite amount of energy, so that
self-ionisation happens for orbits that have achieved a small $k$ and an $L<f(0)$.
The pericentre of the orbit lies at $r_c=(1-\eps)/k^2=L^2/(1+\eps)\approx L^2/2$. 
The critical angular momentum value $L_c=f(0)=0.588057$ is not particularly small, and neither is the critical 
pericentre $r_c=\half f^2(0)=0.172921=23.7\alpha$. 

\begin{figure}[h!]
\label{fig1}
\centerline{ \includegraphics[width=8cm]{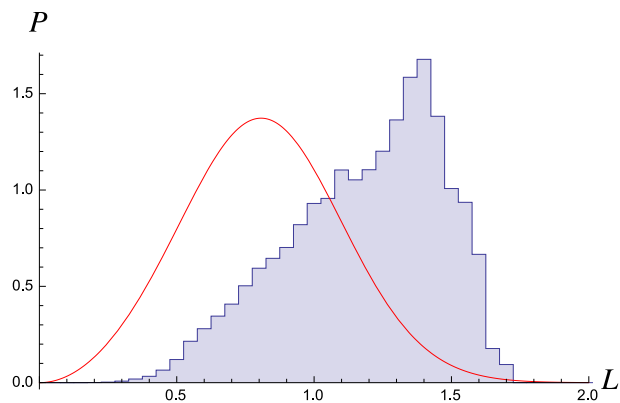}} 
\caption{Distribution of the dimensionless angular momentum $L$ (in units of $\hbar$) from the numerical 
simulation of the SED dynamics reported in ref. \cite{nieuwenhuizen2015simulation1}.
Not much weight lies below $L=0.588$, confirming that when such a value is reached at near-zero energy, 
self-ionisation may occur rapidly and the run is ended. Full curve: the distribution of $L$ from the conjecture for the would-be 
stable ground state distribution,  Eq. (\ref{Ppr}) for $d,\ell_0\to 0$.}
\end{figure}


We are faced with the finding that  the de la Pe\~na-Puthoff stability argument of circular orbits 
\cite{delapena1980introduccion,puthoff1987ground} does not hold for all orbits, 
in particular not for very eccentric ones. This is in full accord with our recent numerical simulations   \cite{nieuwenhuizen2015simulation1}.
Here for short times orbits with reasonable statistics were observed, sometimes going close to the nucleus
or near zero energy, followed by recovery towards less extreme orbits. 
But at some moment no recovery from a near-zero energy orbit was observed, with self-ionisation as a result.
This characteristic is confirmed by the distribution of the dimensionless angular momentum $L$, obtained from these simulations:
In Fig. 1, hardly any weight of orbits occurs lies below $L<0.588$.

\renewcommand{\thesection}{\arabic{section}}
\section{Adding an $1/r^2$ potential}
\setcounter{equation}{0}\setcounter{figure}{0} 
 \renewcommand{\thesection}{\arabic{section}.}

In the hope to find some stable behaviour, we extend the problem without spoiling its analytical solvability.
We consider the presence of a ``radially-directed dipolar force'', stemming from a potential $V_d(r)=-d/2r^2$,
which has the same form as the angular momentum part $L^2/2r^2$ of the kinetic energy.
Such a term may arise from a dipolar force between electron and nucleus, when the deformation of the nucleus,
caused by the electron, is always aligned with the vector between them.

\subsection{Analysing the orbits}

In our classical dynamics approach the unperturbed problem has potential $V$ and Newton force $\vf$
\BEQ\label{withd} 
V(r)=-\frac{1}{r}-\frac{d}{2r^2},\qquad
\vf=-\frac{\vr}{r^3}-d\frac{\vr}{r^4},\qquad
\EEQ
while the random force and the damping lead to the dynamics

\BEQ
\label{rddotnr=}
\ddot\vr=-\frac{\vr}{r^3}-d\frac{\vr}{r^4}-\beta \vE+\beta^2\dddot\vr.
\EEQ

For an unperturbed orbit with energy $\cE=\half\dot\vr^2+V(r)=-\half k^2$ the radial parametrization is still

\BEQ \label{rrada}
r=\frac{1-\eps\cos a}{k^2},\qquad k^3t=\tau_a\equiv a-\eps\sin a,\qquad 
\EEQ
coded by its eccentricity $\eps\equiv\sqrt{1-\kappa^2}$. The presence of $d$ in (\ref{withd}) now leads to the connections
\BEQ
\lambda\equiv kL,\qquad 
\lambda^2= \kappa^2+\delta,\qquad 
\eps^2=1-\lambda^2+\delta, 
\qquad 
\delta=dk^2
\EEQ
so that now $\lambda\neq\kappa$.
 The range of $\eps$ and $\kappa$ is, due to the shape of (\ref{rrada}), at most the interval (0,1).
In repulsive case $d<0$, the physical ranges are $\sqrt{|\delta|}\le\kappa\le1$,  and both $\eps$ and $\lambda$ range from 0 to $\sqrt{1-|\delta|}$; 
then the condition $|\delta|<1$ implies that the orbital energy is bounded from below, viz. $\cE=- k^2/2>-1/2|\delta|$.

For the attractive dipolar force, $d>0$,  $\eps$ and $\kappa$ indeed range between $0$ and $1$, while $\lambda$ takes values between 
$\sqrt{\delta}$ and $\sqrt{1+\delta}$. However, now there is a second family of orbits having $L<\sqrt{d}$, where $L_\eff^2<0$ and $\eps>1$ 
(because $\kappa^2<0$). While such orbit are hyperbolic (unbound) for $d=0$, they are bound for $d>0$: 
{\it these low-angular momentum orbits spiral into and and then out of the nucleus} 
and need to be described within a relativistic framework \cite{boyer2004unfamiliar}.
This property may explain the Darwin term (a $\delta^{(3)}(\vr)$ term) in the relativistic corrections to the H ground state, 
related to our problem with $d=\alpha^2>0$, as an effect of central spiralling.

The relation $\dot\phi=L/r^2$ now brings for orbits with $L_\eff>0$

\BEQ \label{phiad}
\phi_a 
=2\mu\arctan\sqrt{\frac{1+\eps}{1-\eps}}\tan\frac{a}{2},\qquad \mu\equiv \frac{\lambda}{\kappa}=\frac{L}{L_\eff}=\frac{L}{\sqrt{L^2-d}}.
\EEQ
One period in $a$ still takes a time $P=2\pi/k^3$, but it involves  $\mu$ turns in $\phi$, where $\mu$ is in general non-integer.
For $d>0$ each $\phi$-turn takes the shorter time $2\pi /\mu k^3$, so though the solution $r(t)$ remains, $\vr(t)$ rotates faster.
For $d<0$ it rotates slower.
For $\mu\neq 1$,  $\sin(\phi_a-\phi_b)$ and $\cos(\phi_a-\phi_b)$ do not reduce to compact forms like (\ref{sinxycosxy}).

With $\vf^2=(1/r^2+d/r^3)^2$, the rate of energy loss from Eq. (\ref{Eradgen})  generalises to 

\BEQ\label{Elossd}
\langle\dot\cE_{\rm rad}\rangle
=-\beta^2\left\langle\frac{1}{r^4}+\frac{2d}{r^5}+\frac{d^2}{r^6}\right\rangle
=-\beta^2k^8\left(\frac{3 -\kappa^2}{2 \kappa^5}+dk^{2}\frac{5 - 3 \kappa^2}{ \kappa^7}+d^2k^{4}\frac{35 - 30 \kappa^2 + 3 \kappa^4}{8 \kappa^9}\right).
\EEQ
The result is obtained using the properties (\ref{rrada}) of the orbit.

If $d>0$ the orbits with $L<\sqrt{d}$ the orbits spiral into and out of the centre. They are described by the same solution, 
after adding absolute values to keep $\d\phi/\d a$ and $\d t/\d a$ positive,

\BEQ
k^2r_a=|\rho_a|,\qquad \rho_a\equiv 1-\eps\cos a,\qquad k^3\frac{\d t}{d a}=|\rho_a|,\qquad \dot\phi =\frac{k^4L}{\rho_a^2},\qquad
\frac{\d\phi}{\d a} =\frac{kL}{|\rho_a|}.
\EEQ

\subsection{Perturbations of the orbit}

The equations for perturbations $\ddot\vr_1+\nabla\nabla V\cdot\vr_1=\vE$ now read on the comoving  basis

\BEQ\label{r1r2eq}
\ddot r_{1}- r_{1}\dot\phi^2-2\dot r_{2}\frac{L}{r^2}+2\frac{\dot rL}{r^3}r_{2}
-(\frac{2}{r^3}+\frac{3d}{r^4})r_1
&=&E_1 \nn\\
\ddot r_{2}- r_{2}\dot\phi^2+2\dot r_{1}\dot\phi+\ddot\phi r_{1}+(\frac{1}{r^3}+\frac{d}{r^4})r_2&=&E_2 \\
\ddot z_1+(\frac{1}{r^3}+\frac{d}{r^4})z_1&=&E_3  \nn
\EEQ
The homogeneous solutions are still explicit,

\BEQ
\vh^{(1)}(t)&=&\frac{1}{\rho_a}\left(\begin{array}{c}  \eps\sin a  \\  \kappa\mu \\0  \end{array}\right),  \nn \qquad  
\vh^{(2)}(t)=\frac{1}{\rho_a}\left(\begin{array}{c} \kappa^2( \cos a -\eps) - 2\eps\rho_a^2 \\ -\kappa\mu (\kappa^2+\rho_a)\sin a  \\ 0 \end{array}\right) 
+3\eps\tau_a\vh^{(1)},  \qquad \\
\vh^{(3)}(t)&=&\rho_a\left(\begin{array}{c} 0 \\  1   \\ 0 \end{array}\right),  \qquad  \qquad \,
\vh^{(4)}(t)=\frac{\mu}{\rho_a}\left(\begin{array}{c} \kappa(\eps-\cos a)  \\ 
\mu(\kappa^2+\rho_a)\sin a \\ 0  \end{array}\right)+
\eps\frac{\mu^2-1}{\kappa \mu}\phi_a\vh^{(3)}(t),  
\nn\\
\vh^{(5)}(t)&=&\rho_a\sin\phi_a \left(\begin{array}{c}0 \\ 0 \\1\end{array}\right), \qquad \!
\vh^{(6)}(t)=\frac{\eps}{\kappa\mu}\rho_a\cos\phi_a\left(\begin{array}{c}0 \\ 0 \\1\end{array}\right),\qquad 
\EEQ
where also $\vh^{(4)}$ picks up a secular term. Indeed, $\tau_a=k^3t$ and $\phi_a$ increase indefinitely with time,

\BEQ
\tau_a=a-\eps \cos a, \qquad \phi_a =2\mu\arctan\frac{1+\eps}{\kappa}\tan\frac{a}{2}
=\mu(\psi_a+ \tau_a),\qquad 
\EEQ
with  $\psi_a\equiv \phi_a/\mu-\tau_a$ a periodic function in $a$. 

The sum rule for the arctangent,
$\arctan u-\arctan v=\arctan({u-v})/({1+uv})$, implies

\BEQ
\phi_{ab}\equiv\phi_a-\phi_b=2\mu \arctan\frac{\kappa\sin\half(a-b)}{\cos\half(a-b)-\eps\cos\half(a+b)}.
\EEQ
It decomposes as $\phi_{ab}=\mu \tau_{ab}+\psi_{ab}$ with $\tau_{ab}=\tau_a-\tau_b=k^3(t-s)$ and $\psi_{ab}=\psi_a-\psi_b$ periodic.

The solution (\ref{r1G})  and (\ref{Gam=}) remains valid with transverse sector

\BEQ
G_{33}=\Gamma_{33}
=\frac{\rho_a\rho_b}{\kappa\mu k^3}\sin \phi_{ab}
=\frac{(1-\eps\cos a)(1-\eps\cos b)}{\kappa \mu k^3}\sin \phi_{ab}.
\EEQ

\subsection{Statistical rate of energy change from the stochastic field}

For the field contribution to the energy change we have to inspect the behaviour for $s\to t$. While for $d=0$ the problematic terms
cancel, the problem is real for $d\neq 0$. The terms now add up to

\BEQ \label{greeks}
\dot g(t,s)=\frac{d(t-s)^2}{2r^4(t)}+\frac{2 d\eps\sin a}{3kr^6(t)}(t-s)^3
+{\cal O}[(t-s)^4].
\EEQ
To have a well defined $\tau_c\to 0$ limit, this should be regularised. Technically it can be done by adding at fixed $t$ a contribution to the $s$-integral, 
which itself does not contribute to $\langle\dot\cE\rangle$ at finite $\tau_c$:

\BEQ \label{dg2nd}
g \to g_{\rm reg}=g+g_{\rm sub},\qquad  g_{\rm sub}(t,s)=-\frac{d(t-s)^2}{2r^4(t)}.
\EEQ
For the subtraction of the next order  $\,{\cal O}[(t-s)^3]$ term, we point out 
that it yields a $\sin a\,\log\tau_c$ divergency, which vanishes when integrating over a full orbit.

\subsection{Circular orbits}

At $\eps=0$ the cubic term of (\ref{greeks}) vanishes, so after accounting for (\ref{dg2nd}) the problem is regular.
For $d<0$ the maximal $k$ is $k_m=1/\sqrt{|d|}$. The total energy loss rate is

\BEQ
\langle\dot\cE_{\rm tot}\rangle=\half \beta^2k^9
(1 - 6 \lambda^2 + 5 \lambda^3 + 9 \lambda^4 - 12 \lambda^5 + 4 \lambda^6)-\beta^2k^8\lambda^4, 
\qquad \lambda=\sqrt{1+dk^2},
\EEQ
which prevents to pass below the lowest energy coded by $k=k_m$ and $\lambda=0$. For small $k$ and hence $\lambda\approx1$,
the loss term dominates, exhibiting Puthoff-stability.

For $d>0$
\BEQ \label{cdcircorb}
\langle\dot\cE_{\rm tot}\rangle=\half \beta^2k^9 (1 + dk^2)^{3/2}-\beta^2k^8(1+dk^2)^2=\beta^2\frac{k^8}{2}(1+dk^2)^{3/2}\frac{(1-4d)k^2-4}{k+2\sqrt{1+dk^2}}. 
\EEQ
At small $k$ this always shows the stability $\langle\dot\cE_{\rm tot}\rangle<0$. 
For large-$k$ the stability condition $\langle\dot\cE_{\rm tot}\rangle>0$ demands that $d<1/4$,
a condition met curiously already in the quantum approach. So the dipole force does not essentially modify the stability of circular orbits.

In conclusion,  circular orbits retain their de la Pe{\~n}a-Puthoff stability in the presence of the $-d/2r^2$ potential.

\subsection{Very eccentric orbits}

\newcommand{\tot}{{\rm tot}}

The $d=0$ case (pure hydrogen problem) taught us that the remaining interest lies in the limit of very eccentric orbits with energy close to zero, 
where the possibility of self-ionisation looms.  In the limit $\eps\to1,\kappa\to0$ and small $\cE=-\half k^2$, we set $k=\bar k\kappa$, $\lambda=\mu \kappa$,
with the relation $\mu^2=1+d\bar k^2$ from Eqs. (\ref{rrada}), (\ref{phiad}).
Expressed in angular momentum $L$, Eq. (\ref{phiad}) gave $\mu=L/\sqrt{L^2-d}=L/L_\eff$, 
so that  $\bar k=1/\sqrt{L^2-d}=1/L_\eff$. 

As usual, the easy part is the loss term, which can be taken directly from (\ref{Elossd}). It scales as $k^3\bar k^5$,
which implies in the limit $k\to 0$ a finite contribution per period $P=2\pi/k^3$, as before,

\BEQ
\Delta\cE_{\rm rad}=-2\pi \beta^2\bar k^5
\frac{12+40d\bar k^2+35d^2\bar k^4}{8}
=-2\pi \beta^2\bar k^5\frac{7 - 30 \mu^2 + 35 \mu^4}{8}.
\EEQ

The energy gain from the field during one period is

\BEQ
 \Delta \cE_{\rm field}
 =\frac{6\beta^2}{\pi}\int_{-P/2}^{P/2}\d t\int_\minfty^t\d s\frac{\dot g(t,s)}{(t-s)^4}
 =\frac{6\beta^2k^6}{\pi}\int_{-\pi}^\pi \d a  \int_\minfty^a\d b\, \frac{\rho_a\rho_b \dot g(a,b)}{\tau_{ab}^4}.
 \EEQ
We scale  $a\to\kappa x$, $b\to\kappa y$. 
Similar to Eq. (\ref{DeltaEd0}) of the case $d=0$, there results per period a finite limit of the energy gain,

\BEQ
 \Delta \cE_{\rm field}
 =2\pi \beta^2 \bar k^6 G(\mu),\qquad 
 G(\mu)=\frac{3}{\pi^2}\int_\minfty^\infty\d x\int_\minfty^x\d y h(x,y)
\EEQ
Even in this scaling limit this involves a lengthy expression, 

\BEQ
h(x,y)=
\frac{1}{k(x,y)}[h_c(x,y)\cos \phi(x,y)
+h_s(x,y)\frac{\sin\phi(x,y)}{\mu}+h_0(x,y)]+h_{\rm reg}(x,y),
\EEQ
with $\phi(x,y)=2\mu(\arctan x-\arctan y)=2\mu\arctan\frac{x-y}{1+xy}$ and numerators

\BEQ
h_c&=& 15+y^6 (2 \mu^2+x^2-1)+5 y^4 \{x^2 [-2 \mu^2 (x^2+2)+2 x^2+5]+1\} 
\nn\\&+& 10 \mu^2 x (x^2+1)^2 y^3+5 y^2 \{x^2 [-6 \mu^2 (x^2+2)+4 x^2+11]+1\} 
\\&+& 2 x y [\mu^2 (19 x^4+30 x^2-5)+2 (x^6+4 x^4+5 x^2+10)] 
\nn \\ &+& 5 x^2 [-2 \mu^2 (x^4+x^2-1)+2 x^2+3]
-10 (\mu^2-1) x (x^2+1)^2 (y^2+1)^2 \arctan\frac{x-y}{1+x y}, \nn
\\
 h_s&=&4 \mu ^2 x^7+10 \mu ^2 x^6 y+x^5 [8 \mu ^4+\mu ^2 (26-10 y^2)+5 (1+y^2)^2] \nn\\
&+&10 x^3 [-2 \mu ^2 (-2+y^2)+(1+y^2)^2+\mu ^4 (1+y^2)^2]-10 \mu ^2 x^4 y [-1+\mu ^2 (3+y^2)] \nn\\
&-&2 \mu ^2 x^2 y [-5+2 y^4+10 \mu ^2 (3+y^2)]+5 x [(1+y^2)^2-2 \mu ^2 (-3+5 y^2+2 y^4) \\
&+&2 \mu ^4 (-1+6 y^2+3 y^4)]-2 \mu ^2 y [15-2 y^4+\mu ^2 (-5+5 y^2+4 y^4)]\nn\\
&+&10 \mu ^2 (\mu ^2-1) (1+x^2)^2 (1+y^2)^2 \arctan\frac{x-y}{1+x y}, \nn 
\\
h_0&=&
\frac{5 \left(y^2+1\right)}{9 \left(x^2+1\right)}\{-2 \mu ^2 x^6-12 \mu ^2 x^4+4 \left(\mu ^2-1\right) x^3 y \left(y^2+3\right)-18 \mu ^2 x^2   \\&+&
 12 \left(\mu ^2-1\right) x y \left(y^2+3\right)-2 \left(\mu ^2-1\right) y^2 \left(y^2+3\right)^2+2 x^6+12 x^4+18 x^2-27 \left(x^2+1\right)^4\} . \nn
\EEQ
and the common denominator

\BEQ
k(x,y)=\frac{5 }{2^23^4}
(1 + x^2)^2 [3 x + x^3 -( 3y  + y^3)]^4
\EEQ

For the regulator needed to cancel the $(t-s)^3$ term in (\ref{greeks}) we may take

\BEQ
h_{\rm reg}(x,y)=\frac{64 (1-\mu^2) x}{3 (1+x^2)^5}\,\frac{1}{(x-y)(x-y+1)}.\EEQ
Indeed, when integrating it over $y$, a logarithmically divergent term appears, but the result vanishes over a full period, i.e., upon integration over $x$.
Both gain and loss terms being proportional to $k^3$ implies that per period $\pi\le a\le \pi$ there appears, averaged over disorder and over a period,  an  energy change

\BEQ
\Delta\cE=2\pi \beta^2\left [\bar k^6 G(\mu)-\bar k^5\frac{12+40d\bar k^2+35d^2\bar k^4}{8}\right].
\EEQ

Inserting $\bar k=\sqrt{(\mu^2-1)/d}$, this becomes a function of $\mu$  alone,

\BEQ \label{DeltaEmu}
\Delta\cE= 2\pi \beta^2|\mu^2-1|^{5/2}\frac{7 - 30 \mu^2 + 35 \mu^4}{8|d|^3}
\left[H(\mu)- \sqrt{|d|} \,\, \right], \qquad H(\mu)=\frac{8\sqrt{|\mu^2-1|}\, G(\mu)}{7 - 30 \mu^2 + 35 \mu^4}
\EEQ

Self-ionisation is likely prevented when $H(\mu)<\sqrt{|d|}$ for all orbits, that is, for all relevant $\mu$.
In the repulsive case $d<0$ the range for $\mu$ is $0<\mu<1$. Since $H(0)=5.99$ and $H(1)=0$, our statistical argument suggest {\it a  stable bound state} 
for $d<d_c=-35.8$.
This $d_c$ is finite, though rather large.

For $d>0$ the physical domain is  $\mu^2\ge 1$ and $\mu^2<0$. For $\mu^2\ge 1$
we find that $H(\mu)$ is an increasing function. Its asymptotic behaviour can be analyzed. 
The limit $\mu\to\infty$ describes the orbits with lowest possible angular momentum $L_\eff=0$, $L=\sqrt{d}$.
We can scale $y-x\to w(1+x^2)/2\mu$. 
(The leading and subleading terms in $1/\mu$ can just be evaluated; for the second order correction a regularisation is needed, 
a subtraction of total derivatives of the form $\d[w+(w^3+w)\cos w+(w^2+1)\sin w]/\d w$, with coefficients that depend on $u$.) This ends up with

\BEQ
G(\mu)=\frac{35}{16}\mu^3-\frac{15}{8}\mu+{\cal O}(\frac{1}{\mu}),\qquad
H(\mu)=\half -\frac{1}{\mu^2}+{\cal O}(\frac{1}{\mu^4}).
\EEQ
With this shape of $H(\mu)$, Eq. (\ref{DeltaEmu})  predicts that these orbits remain stable only for the extremal value $d_c=\frac{1}{4}$, but then
(\ref{cdcircorb}) predicts that spherical orbits sink to the centre, so no stable cases are found for $d>0$.
The role of orbits spiralling into and out of the centre (the regime $\mu^2<0$)  is left as an open question.

\subsection{Quantum mechanics}

In a quantum approach one would introduce the angular momentum operator $\hat L^2$ and an effective one, $\hat L_{\rm eff}^2=\hat L^2-d$, 
the latter taking  the eigenvalues $l(l+1)-d\equiv \ell_l(\ell_l+1)$ for $l=0,1,2,\cdots$, so that
 $ \ell_l=-\frac{1}{2}+\half \sqrt{(2l+1)^2-4d}$. For the ground state the value $ \ell_0=-\frac{1}{2}+\half \sqrt{1-4d}$ imposes that $d\le \frac{1}{4}$.

 The nonrelativistic Schr\"odinger equation for the radial wave function $\psi(r)$ reads

\BEQ 
-\half \psi''-\frac{1}{r}\psi'+\frac{l(l+1)}{2r^2}\psi-\frac{1}{r}\psi-\frac{d}{2r^2}\psi =\cE\psi.
\EEQ
For  $d=\alpha^2Z^2$ it has an analogy with the Schr\"odinger equation for the spinless relativistic electron in an H-atom
\cite{nieuwenhuizen2005classical}. Indeed, it produces the Dirac square-root formula for the eigenenergies, be it that the total angular momentum operator $\vJ=\vL+\vS$ 
for the spinning electron is reduced to $\vL$ and hence its eigenvalues $j\to l$. The ground state is

\BEQ \label{psi0r=}
 \psi_0(r)&=&\frac{2^{\ell_0}(1+\ell_0)^{-2-\ell_0}}{\sqrt{\pi \Gamma(2+2\ell_0)\,}}r^{\ell_0}e^{-r/(1+\ell_0)},\qquad \cE_0=-\frac{1}{2(1+\ell_0)^2}=-\frac{2}{(1+\sqrt{1-4d}\,)^2}. 
\EEQ
It is normalised according to $4\pi \int_0^\infty\d r\,r^2\psi_0^2(r)=1$.
For the relativistic ``Klein Gordon'' H atom, $d=\alpha^2 Z^2$ affects only the relativistic corrections; 
here we shall keep $d$ as a parameter of order unity, positive or negative.

\subsection{Classical phase space density}

If a stationary ground state exists in the SED problem, it should be expressable as a function of the conserved quantities $\cE$ and $L$,
more precisely, as functions of $R\equiv -1/\cE$ and $L_\eff\equiv \sqrt{L^2-d}$.
Since this task was worked out by us for the ground state and excited states of the relativistic hydrogen atom \cite{nieuwenhuizen2005classical}, 
we can adjust the approach here. For (\ref{psi0r=}) we  coin the shape

\BEQ \label{Ppr}
P_{\vp\vr}(\vr,\vp)=f(L_\eff,R)\equiv C (L_\eff^2R)^{2\ell_0}\,L_\eff R^3e^{-2R/(1+\ell_0)},
\EEQ
Let us verify this and fix the normalisation $C$; the result will be given in (\ref{Cnorm=}). 
The value of the Hamiltonian

\BEQ
H=\half p_r^2+\frac{L_\eff^2}{2r^2}-\frac{1}{r}=\cE=-\frac{1}{R}
\EEQ
 allows to denote the radial velocity $p_r$ and the effective angular momentum $L_\eff=\sqrt{L^2-d}$ as 

\BEQ
p_r=\sqrt{\frac{2(R-r)}{rR}}\,\cos\mu,\qquad L_\eff=\sqrt{\frac{2r(R-r)}{R}}\,\sin\mu,\qquad (0\le \mu\le\pi).
\EEQ
The fact that $L_\eff$ rather than $L$ itself enters here will imply its presence in (\ref{Ppr}).
(The angle $\mu$ is this section should not be confused with the short hand $\mu=\lambda/\kappa$ in the remainder of the text.)
As this implies $\d p_r \d L_\eff=\d R \,\d\mu\,(r/R^2)$, the volume element in momentum space, with
 $0\le\nu\le2\pi$ the azimuthal angle, reads

\BEQ
\d V_p=\d p_r\d p_\perp\d\nu\,p_\perp=\d p_r\d L\d\nu\,\frac{L}{r^2}
=\d p_r\d L_\eff\d\nu\,\frac{L_\eff}{r^2}=\d R\d \mu \d\nu\,\frac{L_\eff}{rR^2}.
\EEQ
Using $L_\eff^2R=2r(R-r)\sin^2\mu$, we have

\BEQ
\int\d V_p\,P_{\vp\vr}(\vr,\vp)&=&2\pi  Cr^{2\ell_0} \int_0^\pi\d\mu\int_r^\infty \d R\, [2(R-r)\sin^2\mu]^{1+2\ell_0}e^{-2R/\ell_0} \nn \\
&=&C\pi^{3/2}(1+\ell_0)^{2+2\ell_0}\Gamma(\frac{3}{2}+2\ell_0) \,r^{2\ell_0}e^{-2r/\ell_0}.
\EEQ
Comparing with $\psi_0^2(r)$ from (\ref{psi0r=}) we see that it has the proper shape, with 

\BEQ \label{Cnorm=}
C=\frac{2^{2+6\ell_0}}{\pi^3(1+\ell_0)^{6+4\ell_0}\Gamma(3+4\ell_0)}.
\EEQ

Another quantity of interest is the distribution function of the conserved quantities $\cE$ and $L$ or $L_\eff$.
As worked out in Eqs. (47) -- (49) of Ref. \cite{nieuwenhuizen2015simulation1}, the relation between  $P_{\cE L_\eff}$ and $P_{\vp\vr}=f(\cE, L_\eff)$ is

\BEQ
P_{\cE L_\eff}(\cE,L_\eff)
&\equiv& \langle\delta(\underline{\cE}(\vp,\vr)-\cE)\delta(\underline{L_\eff}(\vp,\vr)-L_\eff)\rangle
=\frac{2^{5/2}\pi^3 L_\eff}{|\cE|^{3/2}}f(\cE, L_\eff) \nn\\
&=&  \frac{2^{9/2+6\ell_0}}{(1+\ell_0)^{6+4\ell_0}\Gamma(3+4\ell_0)}
\left (\frac{L_\eff^2}{|\cE|} \right)^{2\ell_0}\frac{L_\eff^2}{|\cE|^{9/2}}e^{-2/(1+\ell_0)|\cE|}.
\EEQ
With $\kappa=L_\eff/L_\eff^{\rm max}=kL_\eff$, this becomes

\BEQ
P_{\cE \kappa}(\cE,\kappa)=\frac{\d L_\eff}{\d \kappa}P_{\cE L_\eff}(\cE,L_\eff)= 
\frac{2^{3+4\ell_0}}{(1+\ell_0)^{6+4\ell_0}\Gamma(3+4\ell_0)}
\frac{\kappa^{2+4\ell_0}}{|\cE|^{6+4\ell_0}}e^{-2/(1+\ell_0)|\cE|},
\EEQ
which is normalised to unity, when taking all $0<\kappa<1$ and $0<R<\infty$.

Despite the setback for the SED program for $d=0$ and all $d>0$, it would be interesting to test this distribution for the
regime $d<d_c=-35.8$ where a stable ground state of the problem should occur.

\renewcommand{\thesection}{\arabic{section}}
\section{Discussion}
\setcounter{equation}{0}\setcounter{figure}{0} 
 \renewcommand{\thesection}{\arabic{section}.}

It was put forward by de la Pe\~na 1980 and Puthoff 1987 that circular orbits lead to stability in the hydrogen problem of  Stochastic Electrodynamics 
(SED) \cite{delapena1980introduccion,puthoff1987ground,puthoff2012quantum}.
This gave hope for stability of the full problem, supported by the 2003 Cole-Zou numerical simulations of the dynamics \cite{cole2003quantum}.
Our own, recent simulations improved on these, benefitting from new computer power, new architectures and analytical tricks.
However, it was found that self-ionisation always occurs  \cite{nieuwenhuizen2015simulation1}. This is explained by the present analytical approach.

We have developed the de la Pe\~na-Puthoff statistical theory, where the average energy gain from the field per period of the orbit is evaluated
and compared with the average energy lost by radiation. Since both parts are defined by the unperturbed problem,
the derivation is elegant and prone for study by students.

Our approach predicts indeed that with the electron and the nucleus modelled as point charges, self-ionisation takes place in 
the SED description of the hydrogen atom. Technically, the problem arises from orbits with nearly vanishing energy and moderately small 
angular  momentum, below $L=0.5880\hbar$  (we restore physical units).
The perihelion then lies at distance $r_{\rm min}=\half (L/\hbar)^2a_0=0.173a_0$ from the nucleus. 
The speed takes here its maximum of $0.024c$, so the problem likely is insensitive to relativistic corrections,
as we verified numerically  \cite{nieuwenhuizen2015simulation2}.
Indeed, when one looks at Fig. 1, the problem is not only the self-ionisation for $L<0.588$, but moreover
that the whole distribution is different from what one would expect from a conjecture based on the shape of the 
quantum ground state wavefunction. Hence if the SED program can be saved, either the gain or the loss term, or both,
need to have a different shape.

Next we have added an $-d/2r^2$ potential, which for $d=\alpha^2Z^2$ has connections with the relativistic H problem for a spinless electron.
The problem remains exactly solvable. In the repulsive case $d<0$, $|d|>d_c=28.6$ we predict stability, which would be an interesting test for
our numerical approach.
In the attractive situation $d>0$ we always find instability, even without accounting for 
effects from orbits which spiral into and out of the origin in a finite time \cite{boyer2004unfamiliar}.
Those orbits nevertheless offer in principle a connection with the Darwin term, 
a relativistic delta-function correction to the Hamiltonian for the hydrogen problem, 
so that for some scholars the SED theory will retain a magic spell.

Our results support the conclusions reached by a number of authors regarding the failure of the ``old'' approach in SED 
to account for atomic stability:  atomic orbits are not simply classical orbits perturbed by the stochastic field \cite{claverie1982nonrecurrence}.
For a proper functioning of SED,  an intricate equilibrium state of both the electron and the stochastic field seems to be needed 
 \cite{cetto1996quantum,boyer2011any,santos2012stochastic,delapena2014emerging}.

\acknowledgments{Acknowledgments}
It is a pleasure to thank  Erik van Heusden and Matthew Liska for discussion,
and the latter also for allowing publication of figure 1.

\conflictofinterests{Conflict of Interest}
The author declares no conflict of interest.

\section*{Appendix: The positronium problem in SED}

\renewcommand{\thesection}{\Alph{section}.}
\renewcommand{\thesubsection}{\thesection\arabic{subsection}}
\renewcommand{\theequation}{\thesection\arabic{equation}}
\setcounter{section}{1}

According to standard textbooks, a moving point particle with charge $q_2$ creates an electric field

\BEQ
\vE(\vr,t)=\frac{q_2}{4\pi\eps_0}\frac{s}{(\vs\cdot\vu)^3}[(c^2-u^2)\vu+\vu(\va\cdot\vs)-\va(\vu\cdot\vs)]
\EEQ
where $\vs=\vr-\vw(t_r)$ with $\vr$ the field point where $\vE$ is considered, $\vw(t_r)$ the position of the point charge at the retarded time
$t_r=t-s/c$ and  $\vu=c\,\hat\vs-\vv$. Next to $\vw$, also the velocity $\vv=\dot\vw$ and the acceleration $\va=\ddot\vw$ are taken at the retarded time.
The magnetic field is $\vB=\hat\vs\times\vE/c$.

For slightly relativistic systems, the $\vE$ field can be expanded at fixed $\vr$ in a power series in $v/c$,

\BEQ \label{Erel=}
\vE(\vr,t)=\frac{q_2}{4\pi\eps_0} \left\{  \,  \frac{\hat\vs}{s^2} 
 +\frac{[v^2-3(\hat\vs\cdot\vv)^2]\,\hat\vs}{2c^2s^2}  -\frac{\va+(\hat\vs\cdot\va)\hat\vs}{2c^2s} +\frac{2\dot\va}{3c^3}\,\right\}  +\cdots,
\EEQ
where $\vs=\vr-\vw$ and $\hat\vs=\vs/s$, with now $\vw$, $\vv$, $\va$ and $\dot\va$ taken at the non-retarded time $t$.
We used that 

\BEQ
t-t_r=\frac{s}{c} \left[1+ \frac{ \vv\cdot\hat\vs}{c} 
+\frac{v^2+(\vv\cdot\hat\vs)^2-s\,(\va\cdot\hat\vs)}{2c^2}\right]+\cdots .
\EEQ
The $\dot\va$ term in (\ref{Erel=}) represents a field that is constant in space; it will decay only at distances of order 
$a_0/Z\alpha$, an issue relevant only for relativistic corrections to it,  see the analysis in \cite{nieuwenhuizen2015simulation2}.

In SED a pair of charges $q_{1,2}$ with masses $m_{1,2}$ will satisfy the Abraham-Lorentz equations of motion

\BEQ
m_1\ddot\vr_1&=&\frac{q_1q_2}{4\pi\eps_0} \frac{\vr_1-\vr_2}{r_{12}^3}  +q_1\vE_\perp(t)
+\frac{q_1q_2}{6\pi\eps_0} \frac{\dddot\vr_2}{c^3}+\frac{q_1^2}{6\pi\eps_0} \frac{\dddot\vr_1}{c^3} , \nn \\
m_2\ddot\vr_2&=&\frac{q_1q_2}{4\pi\eps_0} \frac{\vr_2-\vr_1}{r_{12}^3}  +q_2\vE_\perp(t)
+\frac{q_1q_2}{6\pi\eps_0} \frac{\dddot\vr_1}{c^3}+\frac{q_2^2}{6\pi\eps_0} \frac{\dddot\vr_2}{c^3} , 
\EEQ
where we dropped the $1/c^2$ terms, because they are small relativistic corrections to the Coulomb force, 
but kept the radiation terms and added the radiation self-terms.
$\vE_\perp(t)$ is the fluctuating SED field in the dipole-approximation where its spatial dependence is neglected.
In terms of the centre of mass coordinate $\vR=(m_1\vr_1+m_2\vr_2)/M$ and the mutual coordinate $\vr=\vr_1-\vr_2$ this reads

\BEQ
M\ddot \vR&=&Q\vE_\perp
+\frac{Q}{6\pi\eps_0c^3} (Q\dddot \vR+\bar q\,\dddot \vr) 
, \qquad  \nn \\
\mu \ddot \vr &=& \frac{q_1q_2}{4\pi\eps_0} \frac{\vr}{r^3}   +\bar q \,\vE_\perp
+\frac{\bar q}{6\pi\eps_0c^3} (Q\dddot \vR+\bar q\,\dddot \vr) ,
\EEQ
with 

\BEQ
Q=q_1+q_2,\qquad 
\bar q=\frac{m_2q_1-m_1q_2}{m_1+m_2}, \qquad
M=m_1+m_2,\qquad
\mu=\frac{m_1m_2}{m_1+m_2}.
\EEQ
For hydrogen and positronium the charges are opposite ($Q=0$, $\bar q=q_1=-e$), so $\vR(t)=$ const.,
which results in an $\vr$-dynamics similar to that of one electron around an alkali ion of charge $q_2=Ze$,
with its non-zero but negligible centre-of-mass motion and its $\bar q\approx-e$, because $m_1/m_2=m_e/Am_N\ll 1$. 

In conclusion, the results of this paper hold also for positronium after replacing $m_e\to\mu=\half m_e$, because the mutual
electric fields contain relativistic corrections comparable to the self-damping terms.

\end{document}